\documentclass[twocolumn,superscriptaddress,showpacs,aps,prb]{revtex4}
\usepackage{makeidx}
\usepackage{bm}
\usepackage{graphics}
\usepackage[dvips]{epsfig}
\newcounter{fig}

\begin{document}
\title{Universal infrared conductivity of graphite}
\author{L.A. Falkovsky}
\affiliation{L.D. Landau Institute for Theoretical Physics, Moscow
117334, Russia} \affiliation{Institute of the High Pressure
Physics, Troitsk 142190, Russia}

\date{\today}      

\begin{abstract}
The conductivity of graphite is analytically evaluated in the
range of 0.1-1.5 eV, where the electron relaxation processes can
be neglected, and the low energy excitations at the "Dirac" points
are most essential. The value of conductivity calculated per one
graphite layer is close to the universal conductivity of graphene.
The features of the conductivity are explained in terms of
singularities of the electron dispersion in graphite.
\end{abstract}
\maketitle

Since the pioneering  experimental investigations of a single
atomic layer of graphite (graphene) \cite{Novo,ZSA} ,  its
properties attract much attention. Among them, the optical
response is of particular interest. Recently the transmittance of
light throw the graphene monolayer has been measured
\cite{Na,Li,Ma}. The transmittance
$$T=1-\pi\alpha$$
was found  to be frequency independent in a broad range of photon
energy. The result of the experiments is remarkable  because it
involves  the fine structure constant $\alpha$.
 It was discovered that  the real part of the optical conductance of
graphene takes the universal value $$G=\frac{e^2}{4\hbar}$$ which
does not depend on any parameters of graphene. This value agrees
perfectly with the calculations \cite{GSC,FV} ignoring the Coulomb
interactions between electrons. The agreement shows that the
poorly screened Coulomb interaction does not play any role in
graphene for infrared photon frequencies \cite{Mi,SS}.

The intermediate place between 2d graphene and 3d semiconductors
belongs to multilayer graphenes \cite{KA} and graphite, which have
a layered structure with the interlayer distance $c_0=3.35 \AA$
much larger than the nearest-neighbor distance $a_0=1.42 \AA$ in
the layer.  In the study of  graphite \cite{KHC}, it was found
that its optical conductivity per one layer  is very closed to the
universal conductivity of graphene and has evident peculiarities.
The analytic calculation of the in-plane optical response of
graphite  done previously  \cite{Pe} has ignored  coupling between
layers and  no peculiarities have appeared for the infrared
region.

  In the present paper, we evaluate analyticaly the
conductance of graphite in the infrared region of the photon
frequencies. It is known that the low energy electron excitations
in graphene can be described very well with the
Slonczewski-Weiss-McClure  theory \cite{SW}.   The  largest
parameter of the theory, $\gamma_0=3.1$ eV  \cite{PP}, describes
the electron dispersion for  in-layer directions ${\bf k}$. If the
photon energy is less than $\gamma_0$, we can use  the linear
expansion of the in-layer hopping term in the Hamiltonian and
introduce  the constant velocity parameter $v=10^8$ cm/s. The
second parameter in the rang is the interlayer hopping $\gamma_1$
of the order of 0.4 eV which is known from experiments on bilayer
graphene \cite {KCM,Ba}. The parameters $\gamma_3$ and $\gamma_4$
give the corrections  of the order of 10\% to the in-layer
velocity $v$. The electron-hole overlap of the order of 0.02 eV is
determined by parameters $\gamma_2$ and $\gamma_5$ (see Fig.
\ref{gr1}). Therefore, for the photon frequencies larger than 0.1
eV, we  can neglect  the terms with $\gamma_2$ and $\gamma_5$.
 Calculating
such the integral property as conductivity in the
region of the infrared frequencies between 0.1 eV and 1.5 eV,  we can, first, neglect  the small parameters of the theory and, second, use
the $k$-expansion of the in-layer  hopping term. Our results have
the evident analytic form.

In this approximation, the effective Hamiltonian   writes
  near the K-G-H lines of the Brillouin zone in the simple form
\begin{equation}
H(\mathbf{k})=\left(
\begin{array}{cccc}
0 \,    & k_{+} \,& \gamma(z)    \, & 0\\
k_{-} \,& 0     \, & 0\,& 0\\
\gamma(z)    \,  &0 \,& 0 \,  &k_{-}\\
0 \,& 0 \,&k_{+} \,&0
\end{array}%
\right) ,  \label{ham}
\end{equation}%
determined  only by two constants. One is $v=10^8 $cm/s  included
in the definition of the in-plane momentum components,
$k_{\pm}=v(\mp ik_x-k_y)$, and another is the inter-layer
interaction $\gamma_1$ involved in the function
$\gamma(z)=2\gamma_1\cos{z}$. The momentum component $z= k_zc_0$
 is limited by the Brillouin half-zone, $0<z<
\pi/2$ in relative units.

The corresponding eigenenergies are
$$\varepsilon_{1,2}=\frac{\gamma(z)}{2}\pm\sqrt{\frac{1}{4}
\gamma^2(z)+k^2},$$
$$\varepsilon_{3,4}=-\frac{\gamma(z)}{2}\pm\sqrt{\frac{1}{4}
\gamma^2(z)+k^2}.$$ On the K-G-H lines, $k=0$, these equations
determine   two bands  $\varepsilon_{1,4}=\pm \gamma(z)$ and two
degenerate (electron and hole) bands with the energy
$\varepsilon_{2,3}=0$. We have to emphasize that this degeneracy
results from $C_{3v}$ symmetry on the K-G-H line.
\begin{figure}[]
\resizebox{.25\textwidth}{!}{\includegraphics{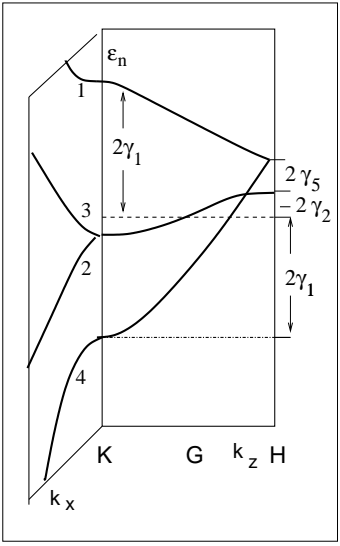}}
\caption{The dispersion of the low energy electron bands in
graphite. } \label{gr1}
\end{figure}

In order to calculate the conductivity, we use the general
expression
\begin{eqnarray}  \label{con}
&&\sigma ^{ij}(\omega) = \frac {2ie^{2}}{(2\pi)^3}\int
d^3k\sum_{k,n\ge m}\left\{ -\frac{df}
{d\varepsilon_n}\frac{v_{n}^{i}v_{n}^{j}}{\omega+i\nu}\right. \\
&&\left. +2\omega  \frac{v_{nm}^{i}v_{mn}^{j}\{f[\varepsilon
_{n}(\mathbf{k})]-f[\varepsilon _{m}(\mathbf{k})]\}}{%
[\varepsilon _{m}(\mathbf{k})-\varepsilon
_{n}(\mathbf{k})]\{(\omega+i\nu)
^{2}-[\varepsilon _{n}(\mathbf{k})-\varepsilon _{m}(\mathbf{k}%
)]^{2}\}}\right\}\,,  \nonumber
\end{eqnarray}
valid in the collisionless limit $\omega\gg\nu$, where $\nu$ is
the collision rate of the carriers,
$f(\varepsilon)=[\exp(\frac{\varepsilon-\mu}{T})-1]^{-1}$ is the
Fermi-Dirac distribution function, and the integral is over the
Brillouin zone.

Here, the first term is the Drude-Boltzmann conductivity
negligible for frequencies larger than the electron-hole overlap.
The second term represents the optical interband transitions of
electrons from the valence 2,4 to conductive 1,3 bands. The real
part of the interband contributions into conductivity arises from
the bypass around the pole at
$\varepsilon_n(\mathbf{k})-\varepsilon_m(\mathbf{k})=\pm\omega$.
The imaginary part is given by the principal value of the
integral.

The  velocity operator
$${\bf v}=\frac{\partial H({\bf
k})}{\partial {\bf k}}$$
 near the K-G-H lines is determined by the Hamiltonian
 (\ref{ham}).
The corresponding  matrix elements  should be calculated in the
representation, where the Hamiltonian has a diagonal form.  The
operator transforming the Hamiltonian to this form can be written
as follows
\[{ U} = \left( \begin{array}{cccc}
 \varepsilon_1/N_1 & \varepsilon_2/N_2 &-\varepsilon_3/N_3 & -\varepsilon_4/N_4 \\
 k_{-}/N_1& k_{-}/N_2 &-k_{-}/N_3 &-k_{-}/N_4 \\
 \varepsilon_1/N_1& \varepsilon_2/N_1 &\varepsilon_3/N_3& \varepsilon_4/N_4\\
 k_{+}/N_1 & k_{+}/N_2& k_{+}/N_3 &k_{+}/N_4
 \end{array} \right)\,, \]
where \(N_{n}^2=2(\varepsilon_n^2+k^2)\)\,. In this
representation, the velocity operator
\[U^{-1}{\bf v}U\]
has the matrix elements
\[\begin{array}{c}
\mathbf{v}_{nn}=\partial \varepsilon_n/\partial {\bf k}\,,\\
\mathbf{v}_{23}=2i(\varepsilon_3-\varepsilon_2)(-k_x{\bf
e}_y+k_y{\bf e}_x)]/N_2N_3\,,\\
\mathbf{v}_{12}=2(\varepsilon_1+\varepsilon_2)(k_x{\bf
e}_x+k_y{\bf e}_y)]/N_1N_2\,,\\
\mathbf{v}_{14}=2i(\varepsilon_4-\varepsilon_1)(-k_x{\bf
e}_y+k_y{\bf e}_x)]/N_1N_4\,,\\
\end{array}\]
where ${\bf e}_i$ are the unit vectors directed along the
coordinate axes. For the real part of conductivity, the
integration in Eq. (\ref{con}) is easily taken at  zero
temperatures $T=0$ in cylindrical coordinates $(k_z, k, \phi)$
over the angle $\phi$ and over $k$ with the help of the
$\delta$-function, $(\omega-x+i\nu)^{-1}\rightarrow
-i\pi\delta(\omega-x)$.
 One obtains
for contributions of the  transitions between the corresponding
valence and conduction bands into the diagonal components of
conductivity (off-diagonal ones equal zero) the following
integrals over $z=k_z/c_0$:

\[\text{Re}~\sigma_{23}=\frac{e^2}{4\pi\hbar
c_0}\int_{0}^{\pi/2}dz\frac{2\gamma(z)+\omega}{\gamma(z)+\omega}\,,\]

\begin{eqnarray}\text{Re}~\sigma_{21}=\frac{e^2}{4\pi\hbar
c_0}\int_0^{\pi/2}dz\frac{\gamma^2(z)}{\omega^2}\theta[\omega-\gamma(z)]\,,
\label{cond}\end{eqnarray}

\[\text{Re}~\sigma_{41}=\frac{e^2}{4\pi\hbar
c_0}\int_0^{\pi/2}dz\frac{2\gamma(z)-\omega}{\gamma(z)-\omega}\theta[\omega-2\gamma(z)]\,,\]
\[\sigma_{43}=\sigma_{21}\,,\]
where $\gamma(z)=2\gamma_1\cos{z}$ and $\theta(x)$ is the step
function.

It is evident from Eqs. (\ref{cond}) (see also Fig. \ref{gr1})
that  the conductivity $\sigma_{23}$ tends to $e^2/4\hbar c_0$ at
the low frequencies $\omega\ll 2\gamma_1$, whereas  other
contributions go to zero in the limit of low frequencies. At
larger frequencies $\omega\gg 2\gamma_1$, the total conductivity
(the sum of $\sigma_{23}$  and $\sigma_{41}$)  tends again to
$e^2/4\hbar c_0$. Therefore, $\sigma_0=e^2/4\hbar c_0$ can be
considered as the universal conductivity of graphite, where
$e^2/4\hbar$ is the conductivity of monolayer graphene and the
factor $1/c_0$ is the number of the layers per the length unit in
the z-direction of graphite.
\begin{figure}[]
\resizebox{.4\textwidth}{!}{\includegraphics{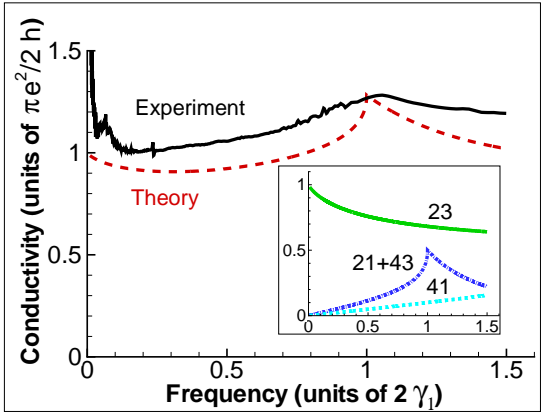}}
\caption{The real part of the graphite conductivity per layer (in
units of $e^2/4\hbar$) versus the frequency (in units of
$2\gamma_1=0.84$ eV); the experimental data \cite{KHC} are shown
in the solid line, results of the present theory in the dashed
line. The insert shows the contributions of  various electron
transitions. } \label{gr1}
\end{figure}
 Integrating in
Eqs. (\ref{cond}), we get finally

\begin{eqnarray}
\text{Re}~ \frac{\sigma_{23}}{\sigma_0}= 1-
\frac{2t}{\pi\sqrt{t^2-1}} \arctan{\sqrt{\frac{t-1} {t+1}}} ,\, t>
1\, ,
 \label{g} \end{eqnarray}
\begin{eqnarray}
\text{Re}~\frac{\sigma_{23}}{\sigma_0}= 1-
 \frac{t}{\pi\sqrt{1-t^2}}
 \ln{\frac{\sqrt{1+t}+\sqrt{1-t}}{\sqrt{1+t}-\sqrt{1-t}}},\,  t< 1\,,
 \nonumber\end{eqnarray}
\begin{eqnarray}
\text{Re}~ \frac{\sigma_{21}}{\sigma_0}= \frac{1}{4t^2 }\left\{
\begin{array}{ll}
 1, & t>1\, , \\
1-\frac{2}{\pi}(\arccos{t} +t\sqrt{1-t^2}) , & t<1\, .
\end{array}
\right. \label{21} \end{eqnarray}
\begin{eqnarray}
\text{Re}~\frac{ \sigma_{41}}{\sigma_0}= 1-
\frac{2t}{\pi\sqrt{t^2-1}} \arctan{\sqrt{\frac{t+1} {t-1}}} ,\, t>
2,\nonumber\end{eqnarray}
\begin{eqnarray}
\text{Re}~ \frac{\sigma_{41}}{\sigma_0} = 1-\frac{2z_1}{\pi}
-\frac{2t}{\pi\sqrt{t^2-1}} \left[
\arctan{\sqrt{\frac{t+1}{t-1}}}\right.\nonumber\\
\left.
-\arctan{\left(\sqrt{\frac{t+1}{t-1}}\tan\frac{z_1}{2}\right)}
\right],\, 1< t < 2\,,\nonumber\\
\text{Re}~\frac{\sigma_{41}}{\sigma_0}= 1-\frac{2z_1}{\pi}+
 \frac{t}{\pi\sqrt{1-t^2}}\left[
 \ln{\frac{\sqrt{1+t}+\sqrt{1-t}}
 {\sqrt{1+t}-\sqrt{1-t}}}\right.\nonumber \\ \left.
 +\ln{\frac{\sqrt{1+t}\tan\frac{z_1}{2}-\sqrt{1-t}}
 {\sqrt{1+t}\tan\frac{z_1}{2}+\sqrt{1-t}}}\right],\, t< 1\, ,
 \nonumber\end{eqnarray}
where $t=\omega/2\gamma_1$ and $z_1=\arccos(t/2)$.

The peculiarity as a kink can be seen in Fig. \ref{gr1}. The
expression (\ref{21}) shows that this kink is located at
$\omega=2\gamma_1$. Taking into account the  kink position
$\omega=0.84$ eV determined experimentally,  the value of
$\gamma_1=0.42$ eV is found in excellent agreement with
experiments on bilayer graphene.

\begin{figure}[]
\resizebox{.4\textwidth}{!}{\includegraphics{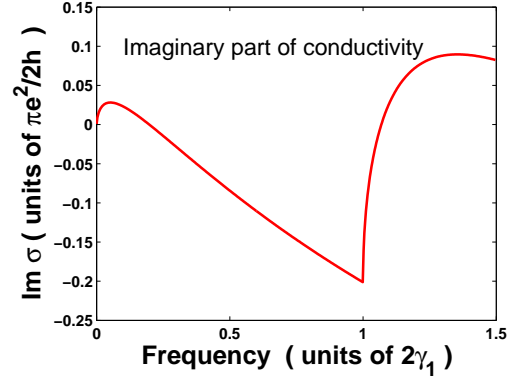}}
\caption{The imaginary part of the graphite conductivity per layer
(in units of $e^2/4\hbar$) versus the frequency (in units of
$2\gamma_1=0.84$ eV). } \label{gr2}
\end{figure}

The contributions of the electron interband transitions into the
imaginary part of conductivity can be integrated  over $k$ at the
zero temperature. The results are obtained in the form of
integrals over $k_z$
 \begin{eqnarray}
\text{Im}~\frac{\sigma_{23}}{\sigma_0}= \frac{2}{\pi^2}
 \int_0^{\pi/2}dz\frac{\omega\gamma(z)}{\gamma^2(z)-\omega^2}
 \ln{[\gamma(z)/\omega]}\, ,
 \nonumber\end{eqnarray}
\begin{eqnarray}
\text{Im}~\frac{\sigma_{21}}{\sigma_0}= \frac{1}{\pi^2}
\int_0^{\pi/2}dz\frac{\gamma(z)}{\omega
}\left(2+\frac{\gamma(z)}{\omega}
 \ln{\frac{|\gamma(z)-\omega|}{\gamma(z)+\omega}}\right)\, ,
 \nonumber\end{eqnarray}
\begin{eqnarray}
\text{Im}~\frac{\sigma_{41}}{\sigma_0}= \frac{1}{\pi^2}
 \int_0^{\pi/2}dz
 \left(\frac{2\gamma(z)-\omega}{\gamma(z)-\omega}
 \ln{|2-\omega/\gamma(z)|}\right.\nonumber\\ \left.-
 \frac{2\gamma(z)+\omega}{\gamma(z)+\omega}
 \ln{(2+\omega/\gamma(z))}\right)\,
 \nonumber\end{eqnarray}
 and shown in Fig. \ref{gr2}. Here, the peculiarity looks like a
threshold at $\omega=2\gamma_1$ and it is more clearly marked in
comparison with the kink in the real conductivity. Both
peculiarities result due to the electron transitions between the
bands $2\rightarrow 1$ and $4 \rightarrow 3$. We should emphasize
that the peculiarities become broader with the temperatures and
the collision processes included.

So far the in-layer conductivity was considered. The estimate of
the inter-layer conductivity can be also done. Since the
conductivity is determined by the ratio  of the corresponding
velocities squared, we have to write
$$v_{z}=\frac{\partial\varepsilon_3}{\partial k_z}\sim
\gamma_1 c_0 \sin(k_zc_0)\,.$$ Then, integrating over $k_z$, we
get
$$\sigma_{z}/\sigma_{0}\sim (\gamma_1 c_0/\hbar
v)^2/2\sim 0.05\,.$$

In conclusion, our calculations reveals that the optical
conductance of graphite can be estimated for frequencies between
0.1 and 1.5 eV multiplying the graphene conductivity $e^2/4\hbar$
by the number of the layers $1/c_0$ per the length unit.  The
Drude-Boltzmann contribution is essential at lower frequencies,
whereas  others interband transitions, e.g. at the M point of the
Brillouin zone contribute into the conductivity  at higher
frequencies. The similar estimate are
applicable for other graphite materials such as nanoribbons. The
kink in the real part of conductivity and the threshold in the
imaginary part appear at the frequency $\omega=2\gamma_1$
determined by the interlayer coupling. The sharpness of the features
are smeared with the relaxation processes and temperatures included.

This work was supported by the Russian Foundation for Basic
Research (grant No. 10-02-00193-a) and by the SCOPES grant IZ73Z0$\_$128026 of the Swiss NSF. The
author is grateful to the Max Planck Institute for the Physics of
Complex Systems for hospitality in Dresden.

\end{document}